\documentclass[preprint,showpacs,preprintnumbers,amsmath,amssymb]{revtex4}
\newcommand{\ds}{\displaystyle}

\usepackage{graphicx}% Include figure files
\usepackage{dcolumn}% Align table columns on decimal point
\usepackage{bm}% bold math

\newcommand{\be}{\begin{equation}}
\newcommand{\en}{\end{equation}}
\newcommand{\bea}{\begin{eqnarray}}
\newcommand{\ena}{\end{eqnarray}}
\begin{document}

%\preprint{GACG/04/2005}

\title{Curvaton reheating in  logamediate inflationary model}

\author{Sergio del Campo}
 \email{sdelcamp@ucv.cl}
\affiliation{ Instituto de F\'{\i}sica, Pontificia Universidad
Cat\'{o}lica de Valpara\'{\i}so, Av. Brasil 2950, Valpara\'{\i}so,
Chile.}

\author{Ram\'on Herrera}
\email{ramon.herrera@ucv.cl} \affiliation{ Instituto de
F\'{\i}sica, Pontificia Universidad Cat\'{o}lica de
Valpara\'{\i}so, Av. Brasil 2950, Valpara\'{\i}so, Chile.}

\author{Joel Saavedra}
\email{joel.saavedra@ucv.cl} \affiliation{ Instituto de
F\'{\i}sica, Pontificia Universidad Cat\'{o}lica de
Valpara\'{\i}so, Av. Brasil 2950, Valpara\'{\i}so, Chile.}

\author{Cuauhtemoc Campuzano}
\email{ccvargas@uv.mx}
\affiliation{Departamento de F\'\i sica, Facultad de F\'\i sica
e Inteligencia Artificial, Universidad Veracruzana, 91000, Xalapa Veracruz,
M\'exico}

\author{Efra\'\i n Rojas}
\email{efrojas@uv.mx}
\affiliation{Departamento de F\'\i sica, Facultad de F\'\i sica
e Inteligencia Artificial, Universidad Veracruzana, 91000, Xalapa Veracruz,
M\'exico}

\date{\today}% It is always \today, today,
             %  but any date may be explicitly specified

\begin{abstract}
In a logamediate inflationary universe model  we introduce the
curvaton field in order to bring this inflationary model to an
end. In this approach we determine the reheating temperature. We
also outline some interesting constraints on the parameters that
describe our models. Thus, we give the parameter space in this
scenario.

\end{abstract}

% PACS, the Physics and Astronomy
                             % Classification Scheme.
%\keywords{Suggested keywords}%Use showkeys class option if keyword
                              %display desired
\maketitle

\section{Introduction}

It is well known that one of the most exciting ideas of
contemporary physics is to explain the origin of the observed
structures in our Universe. It is believed that inflation
\cite{R1} can provide an elegant mechanism to explain the
large-scale structure, as a result of quantum fluctuations in the
early expanding Universe, predicting that small density
perturbations are likely to be generated in the very early
universe with a nearly scale-free spectrum \cite{R2}. This
prediction has been supported by early observational data,
specifically in the detection of temperature fluctuations in the
cosmic microwave background (CMB) by the COBE satellite \cite{R4}.
In this epoch, the predictions of inflation have been detected in
the specific pattern of anisotropies imprinted in the full sky map
of the CMB, as reported, for instance, by the WMAP mission
\cite{R5}. On the other hand, an inflationary-type expansion also
sources a back-ground of primordial gravitational waves \cite{R6},
whose effects still remain undetectable. Forthcoming observations,
such as the PLANCK \cite{R7} or LISA \cite{R8} missions, may
measure effects of relic gravitational waves and offer new trends
for gravitational physics in the near future.

The condition for inflation to occur is that the inflaton field
slow-roll near the top of the potential for sufficiently long
time, so that the vacuum energy drives the inflationary expansion
of the Universe. Many models of inflation have been proposed
\cite{R9,R10}, based on single field or multifield potentials.
Also, they have been constructed in various theoretical schemes.
We distinguish those introduced by Barrow \cite{R11} where the
scale factor $a(t)$ as the asymptotic property that ordinary
differential equations of the form $\ddot{a}=P(a,t)/Q(a,t)$, as $t
\rightarrow\infty$ with polynomials $P$ and $Q$, brings specific
different solutions from which we singularize those named
logamediate inflationary solution. This solution has the
interesting property that the ratio of tensor to scalar
perturbations is small and the power spectrum can be either red or
blue tilted, according to the values of the parameters appearing
in the model \cite{R12}.

 The main motivation to study logamediate inflationary
universe models becomes from the form of the field potential that
appear in this kind of models, i.e.
$V(\phi)\propto\,\phi^{\alpha}\;\exp [-\beta\,\phi^{\gamma}]$.
This potential  includes exponential potential ($\alpha=0$) that
appears in Kaluza-Klein theories, as well as in supergravity, and
in superstring models (see Ref.\cite{P1}). Also, it includes
power-law potentials ($\beta=0$), with models based on dynamical
supersymmetry breaking which motivates  potentials of the type,
$V(\phi)\propto\phi^{-\alpha}$ \cite{P2}. We also find this sort
of potentials  in  models motivated by higher-dimensional
theories, scalar-tensor theories, and supergravity corrections
\cite{P3}. In particular, it was used in Ref.\cite{P4} for
studying inflation with a background dilaton field, and scaling
behavior and other attractorlike solutions were studied in
Ref.\cite{P5}. On the other hand, this potential was used in dark
energy models, driving the observed acceleration of the Universe
at the present epoch \cite{P6}.

In Ref.\cite{A2} was considered  the behavior  of the parameters
$\gamma$ and $\alpha$ in all regions of the  parameter space. When
$0\leq\gamma\leq1$, inflation is generic at late times, for all
values of $\alpha$. For the cases, when $\gamma>1$ and $\alpha\neq
0$, the models are noninflationary.
 When
$\gamma<0$ and $\alpha\geq 0$, inflation comes at early times. For
$\gamma<0$ and $\alpha<0$, inflation occurs at late times and the
inflationary behavior is regulated by the specific value of
$\alpha$. For $\gamma>1$ and $\alpha=0$ the quasi-de Sitter
scenario is manifest at early times, but this is not a
characteristic as $t\rightarrow\infty$, where all models  are
noninflationary. For the case $\gamma>1$ and $\alpha\neq0$, these
models can never inflate.
 In
particular, if $\gamma=0$, the solutions present
polynomial-chaotic or intermediate inflationary behavior depending
upon the sign of $\alpha$, and for the special case when
$\alpha=0$ the de Sitter solution is obtained. The power-law
solutions occur for $\gamma=1$ and $\alpha=0$  and if
$\alpha\neq0$  the behavior will asymptote to the power-law
solution at large times $t$. If $0<\gamma<1$ the scale factor is
proportional to $\exp(\ln^{(2-\gamma)/\gamma})$ as
$t\rightarrow\infty$.

One of the drawbacks of this model rests on the impossibility to
bring inflation to an ends. In fact, at the end of inflation the
energy density of the universe is locked up in a combination of
kinetic and potential energies of the scalar field, which drives
inflation \cite{R9}. One path to defrost the universe after
inflation is known as reheating \cite{R14}. During reheating, most
of the matter and radiation of the universe are created usually
via the decay of the inflaton field leading to a creation of
particles of different kinds, while the temperature grows in many
orders of magnitude. It is at this point where the universe
matches the Big-Bang model. In this process is the particular
interest in in the quantity known as the reheating temperature,
$T_{rh}$. The reheating temperature is related to the temperature
of the Universe when the radiation epoch begins.

The oscillations of the scalar inflaton field are an essential
part for the standard mechanism of reheating. However, there are
some models where the inflaton potential does not have a minimum
and the scalar field does not oscillate. Here, the standard
mechanism of reheating does not work \cite{R15}. These models are
known in the literature like nonoscillating (NO) models
\cite{R16,R17}. The NO models correspond to runaway fields such as
module fields in string theory which are potentially useful for
inflation model-building because they presents flat directions
which survive the famous $\eta$-problem of inflation \cite{R18}.
This problem is related to the fact that between the inflationary
plateau and the quintessential tail there is a difference of over
a 100 orders of magnitude.

%On the other hand, an important use of NO models is quintessential
%inflation, in which the tail of the potential can be responsible
%for the accelerated expansion of the present universe \cite{R19}.
%In this respect, and due to the present phase of accelerated
%expansion, it is required that its minimum has not been reached
%yet, in order for the residual energy density not to be zero at
%present time.**

The first mechanism of reheating in this kind of model was the
gravitational particle production \cite{R20}, but this mechanism
is quite inefficient, since it may lead to certain cosmological
problems \cite{R21,R22}. An alternative mechanism of reheating in
NO models is the instant preheating, which introduce an
interaction between the inflaton scalar field an another scalar
field \cite{R16}. Another possibility for reheating in NO models
is the introduction of the curvaton field, $\sigma$, \cite{R23},
which has recently received a lot of attention in the literature
\cite{R24,R25}. The curvaton approach is an interesting new
proposal for explaining the observed large-scale adiabatic density
perturbations in the context of inflation. Here, the hypothesis is
such that the adiabatic density perturbation originates from the
-"curvaton field"- and not from the inflaton field. In this
scenario, the adiabatic density perturbation is generated only
after inflation, from an initial condition which corresponds to a
purely isocurvature perturbation \cite{R26}.  Following,
Ref.\cite{M1} (see, also Ref.\cite{M2}) we adopt the "curvaton
hypothesis", where the inflaton  perturbation is taken to be less
than 1$\%$ of the observed value. Using the COBE normalization at
the pivot scale, we can set an upper bound for the power spectrum
of inflation, where $P_{\zeta_
{\phi}}^{1/2}\lesssim\,0.01\,P_\zeta^{1/2}\simeq
5\times\,10^{-7}$. Here, $P_{\zeta_ {\phi}}^{1/2}$ and
$P_{\zeta}^{1/2}$ are the power spectrum of the inflaton field and
curvaton field, respectively. On the other hand, generally
inflationary models suggest that inflation took place at energy
comparable to that of grand unification, where the energy scale is
approximately  $V_*^{1/4}\approx 10^{15-16}$ GeV , where $V_*$ is
the effective potential associated with  the inflaton field
evaluated when the cosmological scales exit the horizon
 \cite{M3}.  In the  context of string
 landscape  supersymmetry sets the value of $V_*$ at
 scales  typically much less than the grand unified scale. One
 way to liberate inflation from the constraint given by $V_*^{1/4}\approx 10^{15-16}$ GeV
is to consider that the curvature perturbations generated during
inflation are due to quantum fluctuations of the curvaton field,
in which case $V_*^{1/4}\approx 10^{15-16}$ GeV turns into an
upper bound \cite{M1}. It is assumed  that the curvaton field does
not influence the dynamics of the inflaton field, but becomes
important after inflation has ended, when it imprints its
curvature perturbation onto the universe \cite{R23}. Under this
hypothesis, it is possible to diminish this constraint
substantially \cite{M4}. However, we should note that, one can
have cases in which the fluctuations generated by both, the
inflaton and a curvatonlike field are relevant \cite{M5,M6}.

In the framework of logamediate inflationary universe models we
would like to introduce the curvaton field as a mechanism to bring
logamediate inflation to an end. Therefore, the main aim of this
paper is to carry out the curvaton field into the logamediate
inflationary scenario and see what consequences we may derive. The
outline of the paper goes as follow: in Sec. II we give a brief
description of the logamediate inflationary scenario. In Sec III
the curvaton field is described in the kinetic epoch. Section IV
describes the curvaton decay after its domination. Section V
describe the decay of the curvaton field before it dominates.
Section VI studies the consequences of the gravitational waves. At
the end, Sec. VII includes our conclusions.

\section{Logamediate inflation  Model \label{secti}}

In order to introduce the logamediate inflationary universe model
we start with the corresponding field equations that must satisfy
the scalar field in a flat Friedmann-Robertson-Walker (FRW)
background

\begin{equation}
 3\;H^2=\frac{\dot{\phi}^2}{2}+V(\phi)
 \label{key_02},
\end{equation}
and
\begin{equation} \ddot{\phi}+3H\dot{\phi}=-\frac{\partial
V(\phi)}{\partial\phi},
\label{3}
\end{equation}
where  $H \equiv \dot{a}/a$ is the Hubble factor, $a=a(t)$ is the
scale factor. Here, $\phi$ is the standard inflaton field and
$V(\phi)$ its associated effective scalar potential, the  dots
denote derivative with respect to the cosmological time $t$, and
we shall use  units such that $8\pi G=8\pi/m_p^2=c=\hbar=1$,
\,\,$m_p$ being the Planck mass.

The main assumption in the logamediate inflationary universe model
is that the scale factor $a(t)$ expands by means of the asymptotic
form \cite{R11, R12}
\begin{equation}
a(t)=\exp[\,A\,(\ln t)^{\lambda}],\label{at}
\end{equation}
with $t>1$. Here $A>0$, and $\lambda>1$ are constants. The Hubble
parameter as a function of the cosmological times $t$ becomes
\begin{equation}
H=\frac{\dot{a}}{a}=\frac{A\,\lambda (\ln\,t)^{\lambda-1}}{t},
\end{equation}
and since $A\,\lambda>0$ we get expanding universe. Note that when
$\lambda =1$ this model reduces the well known power-law
inflation, $a\propto t^p$, where $p=A$ with $A>1$. From Eqs.
(\ref{key_02}-\ref{at}), we have  $\dot{H}=-\dot{\phi}^2/2$ %the expressions for the scalar potential
%$V(\phi)$
 and the scalar field $\phi(t)$ result
\begin{equation}
\phi = \phi_0 + \gamma\,(A\lambda)^{1/2} \,(\ln
t)^{\frac{1}{\gamma}},
%\frac{2(A\lambda)^{1/2}}{\lambda + 1}\,(\ln t)^{\frac{\lambda +1}{2}},
%(2\;A\beta\;t^{f})^{1/2},
\label{pat}
\end{equation}
where   {\small $\gamma = \frac{2}{\lambda +1}$}.

Assuming the set of slow-roll conditions, i.e., $\dot{\phi}^2 \ll
V(\phi)$ and $\ddot{\phi}\ll 3H\dot{\phi}$, and setting
$\phi_0=0$, without loss of generality, the scalar potential can
be written as \cite{R12}
\begin{equation}
V(\phi)= V_0 \,\phi^\alpha \,\exp[-\beta\,\phi^\gamma].
%\frac{8A^2}{(\beta+4)^2}\left[\frac{\phi}{(2A\beta)^{1/2}}\right]^{-\beta}
%\left[6-\frac{\beta^2}{\phi^2}\right]
\label{PP}
\end{equation}
Here,  the parameters $\alpha$ and $\beta$   are defined by
{\small $\alpha = \frac{4(\lambda -1)}{(\lambda + 1)}$ and $\beta
=2 \left[ \frac{(\lambda + 1) }{ (2
\sqrt{2}(A\lambda)^{1/2})}\right]^{2/(\lambda + 1)}
%\equiv 4(f^{-1}-1)$.
$}, respectively. Furthermore, we have defined
\begin{equation}
V_0 = \frac{3}{2} (A\lambda)^2 \,\beta^{2(\lambda -1)}.
%3(A\lambda)^2 \left( \frac{\lambda +1}{2\sqrt{2} (A\lambda)^{1/2}}
%\right)^{\frac{4(\lambda -1)}{(\lambda +1)}} .
\end{equation}
Note that this kind of potential does not present a minimum for
large values of the field $\phi$.
% This class of scalar potentials
%were examined in Refs.\cite{A1,A2}.
This potential was originally studied by Barrow \cite{A1} (see
also Ref.\cite{A2}), where it was shown that the condition
$\gamma=2/(\lambda+1)\leq 1$ was needed for inflation  to occurs
at large values of $\phi$.

%When $\gamma=0$ and $\alpha$ is positive we have the chaotic
%inflation. For $\gamma=0$ and $\alpha<0$ the evolution asymptotes
%to  intermediate inflation and when $\gamma=1$ and $\alpha=0$ we
%retrieve the power-law solution for the scale factor\cite{B2}. The
%choice $\gamma=\alpha=0$ corresponds to Guth's model with
%exponential expansion\cite{Gu2}.
%In
% was shown that using asymptotic techniques to
%ascertained the behavior of the system at large scalar field
%$\phi$ in all regions of the $\alpha-\gamma$ parameter space.

\noindent
The Hubble factor as a function of the inflaton field, $\phi$,
becomes
\begin{equation}
H(\phi)= H_0 \, \phi ^{\frac{\alpha}{2}}\, \exp\left[-
\beta\,\phi^\gamma/2 \right],
%A\;f\;(2\;A\beta)^{\beta/4}\;\phi^{-\beta/2}.
\label{HH}
\end{equation}
where   $H_0=\sqrt{V_0/3}$.

The slow-roll parameters, $\varepsilon$ and $\eta$, are defined by
$\varepsilon=\frac{V'^2}{2V^2}$ and $\eta=V''/V$, respectively.
Here the prime denotes derivative with respect to the inflaton
field $\phi$. In the present case they read
\begin{equation}
\varepsilon = \frac{1}{2\phi^2} \left( \alpha - \beta \,\gamma
\phi^\gamma\right)^2,\;\;\;\mbox{and}\;\;\;  \eta= -
\frac{1}{\phi^2} \left[ \alpha + \beta\,\gamma\,(\gamma
-1)\,\phi^\gamma - \frac{1}{2} (\alpha -
\beta\,\gamma\,\phi^\gamma)^2  \right],
\end{equation}
and its ratio results in {\small $\frac{\varepsilon}{\eta}= \left[
1 -2 \frac{\alpha + \beta\,\gamma\, (\gamma
-1)\,\phi^\gamma}{(\alpha - \beta\,\gamma\,
\phi^\gamma)^2}\right]^{-1}$}.

Note that,  the parameters $\varepsilon$ and $\eta$ diverges when
the scalar field $\phi\rightarrow 0$. Also, $\varepsilon$ is
always larger than $\eta$ since $\beta$ is positive ($A>0$ and
$\lambda>1$), then $\varepsilon$ reaches unity before $\eta$ does.
In this way, we may establish that the end of inflation is
governed by the condition $\varepsilon=1$. From the condition
$\varepsilon=1$ we can distinguish two possible solutions for the
scalar field, at the end of inflation: $\phi_e=
\frac{\alpha}{\sqrt{2}}$ for the value of $\alpha
> \beta\,\gamma\,\phi^{\gamma}$ and  {\small $\phi_e =\left(
\frac{\beta^2\,\gamma^2}{2}\right)^{\frac{1}{2(1-\gamma)}} $}, for
$\alpha < \beta\,\gamma\,\phi^{\gamma}$. From now on, the
subscript $e$ will be used to denote the end of the inflationary
period.   The maximum of the potential occurs when the parameter
$\varepsilon=0$ ($\partial V/\partial\phi=0$) and the value of the
scalar field in this maximum of the  potential is
$\phi_{i}=(\alpha/(\beta\gamma))^{1/\gamma}$. In the following, we
study our inflationary scenario and the reheating of the Universe
for values of the scalar field, such that $\phi_i\leq \phi$. Then,
we are taking  {\small $\phi_e =\left(
\frac{\beta^2\,\gamma^2}{2}\right)^{\frac{1}{2(1-\gamma)}} $},
because with this choice it is possible to get a continuous
transition from the inflationary age to the kinetic phase.

\section{The curvaton field  \label{sectii}}

When inflation has finished, the model enters
%to a new period which is called
to the -"kinetic epoch"- (or -"kination"-, for simplicity)
\cite{Ag1}. In this epoch, the term $\ds \frac{\partial
V(\phi)}{\partial \phi}$ is negligible compared to the friction
term $3H\dot{\phi}$ in the field Eq. ({\ref{3}}). Hereafter, with
%In the following we will use
the subscript (or superscript) -"k"- we label different quantities
at the beginning of this epoch. The kinetic epoch does not occur
immediately after inflation; there may exist a middle epoch where
the potential force is negligible with respect to the friction
term \cite{Ag2}.  During the kination epoch we have that
$\dot{\phi}^2/2>V(\phi)$ which could be seen as a stiff fluid
since the relation between the pressure $P_\phi$ and the energy
density $\rho_\phi$, corresponds to the relation
$P_\phi\simeq\rho_\phi$.

During the kinetic epoch we have  $
3\;H^2\,=\rho_\phi\simeq\frac{\dot{\phi}^2}{2} $ and $
\ddot{\phi}+3H\dot{\phi}=0$, where the latter equation could be
solved and gives $\dot{\phi}=\dot{\phi}_k \left ({\frac{
a_k}{a}}\right )^3$. This expression yields
\begin{equation}
\rho_{\phi}(a)=\rho_{\phi}^{k}\left (\frac{a_{k}}{a}\right )^6
\label{rho},
\end{equation}
and the Hubble parameter becomes
\begin{equation}
H(a)=H=H_{k}\left (\frac{a_k}{a}\right )^3
  \label{h},
\end{equation}
where
$H_{k}^2=\frac{\rho_{\phi}^{k}}{3}\simeq\frac{\dot{\phi}_k^2}{6} $
is the value of the Hubble parameter at the beginning of the
kination.

We now study the dynamics of the curvaton field, $\sigma$, through
different stages.  We consider that the curvaton field obeys the
Klein-Gordon equation and, for simplicity, we assume that its
scalar potential associated with this field is given by
$U(\sigma)=\frac{m^2\sigma^2}{2}$, where $m$ is the curvaton mass.
This study allows us to find some constraints on the parameters
and thus, to have a viable curvaton scenario.

First, we assume  that the energy density $\rho_\phi$, associated
with the inflaton field, is the dominant component when it is
compared with the curvaton energy density, $\rho_\sigma$, i.e.,
$\rho_\phi\gg\rho_\sigma$. In the next stage, the curvaton field
oscillates around the minimum of the effective potential
$U(\sigma)$. Its energy density evolves as a nonrelativistic
matter and, during the kinetic epoch, the universe remains
inflaton-dominated. Finally, the last stage corresponds to the
decay of the curvaton field into radiation and then the standard
big-bang cosmology is recovered.

%Eventually, these scalar fields fall into the kinetic epoch
%where the inflaton energy density remains dominating. At this
%stage the curvaton energy density evolves gradually as non-relativistic
%matter being behaved as $\rho_\sigma\propto\,a^{-3}$.
%In the meantime the curvaton field oscillates around the minimum of its
%effective potential $U(\sigma)$.
%In the course of the inflationary regime it is assumed that the  curvaton
%field is effectively massless \cite{dimo,postma,urena,cdch}. In
%the same period the curvaton rolls down its potential until its
%kinetic energy is depleted by the exponential expansion. This
%kinetic energy has almost vanished and becomes frozen.
%The final stage corresponds to the
%decay of the curvaton field into radiation and thus the standard
%Big-Bang cosmology is recovered.
%The curvaton field assumes roughly a constant value, $\sigma_*\approx
%\sigma_e$. Here, $"*"$ refers to the epoch when
%the cosmological scale exits the horizon.

In the inflationary regime it is supposed that the  curvaton field
mass $m$ satisfied the condition $m\ll H_e$, and its dynamics is
described in detail in Refs.\cite{dimo,postma,cdch}. During
inflation, the curvaton would roll down its potential until its
kinetic energy is depleted by the exponential expansion and only
then, i.e. only after its kinetic energy has almost vanished, it
becomes frozen and assumes roughly a constant value, i.e.,
$\sigma_*\approx \sigma_e$. The subscript "$*$" here refers to the
epoch when the cosmological scales exit the horizon.

During the kinetic epoch  the Hubble parameter decreases so that
its value is comparable with the curvaton mass, and the curvaton
mass is of the order of Hubble parameter, i.e.,  $H\simeq m$. Then
from Eq. (\ref{h}), we obtain
\begin{equation}
\frac{m}{H_{k}}=\left (\frac{a_k}{a_m}\right )^3,\label{mh}
\end{equation}
where the subscript $'m'$ stands for quantities evaluated at the
time when the curvaton mass, $m$, is of the order of $H$.

In order to prevent a period of curvaton-driven inflation the
universe must still be dominated by the inflaton field, {\it
i.e.,} $\rho_{\phi}|_{a_m}=\rho_{\phi}^{m}\gg\rho_{\sigma}$. Over
the inflation period the effective potential does not change
substantially, because it is reasonable to suppose that
$\rho_{\phi}^{m}\gg\rho_{\sigma}\sim U(\sigma_e)
\simeq\,U(\sigma_*)$. The quoted inequality allows us to find a
constraint on the values of the curvaton field $\sigma_*$ at the
moment when $H\simeq m$, since
%Eq.(\ref{key_2}) gives
\begin{equation}
\frac{m^2\sigma_*^2}{2\rho_\phi^{m}}=\frac{\sigma_*^2}{6}\ll\,\,1\,,\label{pot}
\end{equation}
%{\underline{which implies that the curvaton field $\sigma_*$ satisfies the
%constraint $\sigma_*^2\ll 6$}, where we have used
or equivalently $\sigma_*^2\ll 6$.

Now, at the end of inflation the ratio between the potential energies
results
\begin{equation}
\frac{U_e}{V_e}=\frac{m^2\sigma_*^2}{6
H_e^2}\ll\,\frac{m^2}{H_e^2} \label{u},
\end{equation}
here, we have used Eq.(\ref{pot}).

Since the  curvaton energy becomes subdominant at the end of
inflation, i.e., $V_e\gg U_e$, then  the curvaton mass should obey
the constraint $m^2 \ll H_e ^2$,   and using the relations $
H^2_e=V_e/3$, Eq.(\ref{u}) and
$\phi_e=(\beta^2\gamma^2/2)^{1/2(1-\gamma)}$, we get
%which may be specified by the constraint
%
\begin{equation}
m^2\ll \frac{V_0}{3}\,  \left( \frac{2}{\beta^2 \gamma^2}
\right)^{\frac{\alpha}{2(\gamma -1)}}\,\exp \left[ -\beta \,\left(
\frac{2}{\beta^2 \gamma^2} \right)^{\frac{\gamma}{2(\gamma
-1)}}\right] .
%\left[\frac{2\,A}{\beta+1}\right]^{\beta/2},
\label{one}
\end{equation}

After the curvaton field becomes effectively massive, its energy
decays as a nonrelativistic matter in the form $ \rho_\sigma=
\frac{m^2\sigma_*^2}{2} \left( \frac{a_m}{a}\right)^3$. In the
following, we will study  the decay of the curvaton field  in two
possible different scenarios.

\section{Curvaton Decay After Domination\label{sectiv}}

 For the first scenario, when  the curvaton field comes to
dominate the cosmic expansion {\it i.e.,} $\rho_\sigma>\rho_\phi$,
there must be a moment in which the inflaton and curvaton energy
densities match. We are going to assume that this  happens when
$a=a_{eq}$. Then, from Eqs.(\ref{rho}) and (\ref{h}), and bearing
in mind that $\rho_\sigma\propto\,a^{-3}$, we get
\begin{equation}
\left.\frac{\rho_\sigma}{\rho_\phi}\right|_{a=a_{eq}}=\frac{m^2\sigma_*^2}{2}\frac{a_m^3\;a_{eq}^3}{a_k^6\;\rho_\phi^k}
=\frac{m^2\sigma_*^2 a_m^3 a_{eq}^3}{6\;H_k^2 \; a_k^6} =
1\label{equili},
\end{equation}
where we have used the relation $3\,H_k^2=\rho_\phi^k$ together
with $H_k\left(\frac{a_k}{a_{eq}}\right)^3=\frac{m\sigma_*^2}{6}$
and Eq.(\ref{mh}).

In terms of the curvaton parameters, the Hubble parameter,
$H(a_{eq})=H_{eq}$ can be rewritten as
\begin{eqnarray}
H_{eq}&=&
H_{k}\left(\frac{a_k}{a_{eq}}\right)^3=\frac{m\;\sigma_*^2}{6},\label{heq}
\end{eqnarray}
where we have considered Eqs.(\ref{h}), (\ref{mh}) and (\ref{equili}).

When the curvaton decays after domination  we require that the
following condition is fulfilled, $\rho_\sigma > \rho_\phi$, in
addition to the decay parameter $\Gamma_\sigma < H_{eq}$. Since
the decay parameter $\Gamma_\sigma$ is constrained by
nucleosynthesis, it is required that the curvaton field decays
before nucleosynthesis, which means $H_{nucl}\sim 10^{-40}<
\Gamma_\sigma$.
%(in units of Planck mass $m_p$).
Hence, the constraint upon the decay parameter is
\begin{equation}
10^{-40}<\Gamma_{\sigma}<\frac{m\;\sigma_*^2}{6} .\label{gamm1}
\end{equation}

The curvaton approach is potentially valuable in the search of
physical constraints on the parameters appearing in the
logamediate expanding model by studying the scalar perturbations
related to the curvaton field $\sigma$. During the time in which
the fluctuations are inside the horizon, they obey the same
differential equation of the inflaton fluctuations. We may
conclude that they acquire the amplitude $\delta\sigma_*\simeq
H_*/2\pi$. On the other hand, outside of the horizon, the
fluctuations obey the same differential equation like  the
unperturbed curvaton field and then, we expect that they remain
constant over inflation. The spectrum of the Bardeen parameter,
$P_\zeta$, whose observed value is  $P_\zeta\simeq 2.4\times
10^{-9}$ \cite{WMAP3}, allows us to determine the value of the
curvaton field, $\sigma$, evaluated at the epoch when the
cosmological scales exit the horizon.  This becomes in terms of
the parameters $A$ and $\beta$. At the time when the decay of the
curvaton field occurs the  parameter $P_\zeta$ results to in
\cite{ref1u}
\begin{equation}
P_\zeta\simeq
\frac{H_*^2}{9\pi^2\sigma_*^2}\simeq\frac{1}{9\pi^2\sigma_*^2}\,H_0^2
\left[ B\,N_* + \,\phi_e ^{\lambda\gamma} \right]^{\frac{2(\lambda
-1)}{\lambda}}\,\exp\left[ -\beta \left( B\,N_* + \phi_e
^{\lambda\gamma} \right)^{1/\lambda} \right]  , \label{pafter}
\end{equation}
where {\small $B=A^{-1}\; \left(  A\lambda\gamma^2
\right)^{\lambda /(\lambda +1)}$}. Here, the number of e-folds,
$N_*$,  is determined by  $N_* = \int_{t_*} ^{t_e} H(t')dt'$.
After a rather involved lengthy but straightforward computation we
get
\begin{equation}
N_* = A \left[ \frac{(\lambda +1)}{2 (A\lambda)^{1/2}}
\right]^{\frac{2\lambda}{\lambda + 1}}\left( \phi_* ^{\frac{2\lambda}{\lambda + 1}}
 - \phi_e ^{\frac{2\lambda}{\lambda + 1}}\right),\label{nn}
\end{equation}
wich was used in determining  Eq.(\ref{pafter}).

%The spectrum of fluctuations is automatically Gaussian for
%$\sigma_*^2\gg H_*^2/4\pi^2$, besides independent of
%$\Gamma_\sigma$ \cite{ref1u}. This feature will simplify the
%analysis in the space parameter in the logamediate model.

%It is worthy mention that Eq. (\ref{pafter}) encodes a transcendental
%equation for $A$ in terms of the parameters $P_\zeta, N_*$ and $\lambda$.
%This comprises an inconvenience in obtaining a closed
%expression for $A$.

%From expression (\ref{pafter}), we may write
%\begin{equation}
%A=\left[\frac{27\pi^2}{48}\;\sigma_*^2\;(\beta+4)^2\;P_{\zeta}\left(\frac{\beta+1}{2}-N_*\right)^{\beta/2}
%\right]^{\frac{2}{\beta+4}}, \label{18}
%\end{equation}
%where
%\begin{equation}
%N_*=\int_{t_*}^{t_e}\;H(t')\,dt'=\frac{1}{2\beta}\;(\phi_e^2-\phi_*^2),
%\end{equation}
%defines  the number of the e-folds corresponding to the
%cosmological scales, i.e. the number of remaining inflationary
%e-folds at the time when the cosmological scale exits the horizon.
%Note that the parameter $\beta$ satisfies $\beta>2\,N_*-1$ or
%equivalently $4\,(2N_*+3)^{-1}>f$.

The constraint given by Eq. (\ref{gamm1}) becomes

\begin{equation}
\Gamma_\sigma < \frac{m\,H_0^2}{54\pi^2\,P_\zeta}\,\left[ B\,N_* +
\,\phi_e ^{\lambda\gamma} \right]^{\frac{2(\lambda
-1)}{\lambda}}\,\exp\left[ -\beta \left( B\,N_* + \phi_e
^{\lambda\gamma} \right)^{1/\lambda} \right],\label{ww}
\end{equation}
which provides an upper limit on $\Gamma_\sigma$ when the curvaton
field decays after domination.

We consider  the the hypothesis that the inflaton field curvature
perturbation is taken to be less than 1$\%$ of the observed value,
i.e. $P_{\zeta_{\phi}}\lesssim\,0.0001\,P_\zeta$, where
$P_{\zeta_{\phi}}$, is given by
$P_{\zeta_{\phi}}=V/(24\pi^2\,\varepsilon)$ \cite{M3}. In this
way, we can set a new constraint for the decay parameter
$\Gamma_\sigma$ given by
\begin{eqnarray}
\Gamma_\sigma<\,\frac{2}{3}\;\times10^{-4}\,\beta^2\,\gamma^2\;\;m\;\;
\left[\frac{N_*}{A}\,\left(\frac{2(A\lambda)^{1/2}}{\lambda+1}\right)^{(\lambda+1)/2\lambda}+
\left(\frac{\beta\gamma}{\sqrt{2}}\right)^{(\lambda-1)/\lambda}\right]^{(1-\lambda)/\lambda}\,\,.\label{ew}
\end{eqnarray}
Here, we have used Eqs. (\ref{gamm1}), (\ref{pafter})  and
(\ref{nn}).

Now we turn to give  the constraints on the parameters $A$ and
$\beta$ by using the big bang nucleosynthesis (BBN) temperature
$T_{BBN}$. We know that reheating occurs before the BBN where the
temperature is of the order of $T_{BBN}\sim 10^{-22}$,  and thus
the reheating temperature should satisfy    $T_{reh}>T_{BBN}$. By
using that $T_{reh}\sim \Gamma_\sigma^{1/2}\,>\,T_{BBN}$ we obtain
a new constraint
\begin{equation}
H_{*}^2=\frac{V_*}{3}=\frac{V_0}{3}\,\phi_*^\alpha\,e^{-\beta\phi_*^\gamma}
%=16 \left( \frac{A}{4+\beta}\right)^2 \left
%[\frac{2\,A}{\beta+1-2\,N_*} \right]^{\beta/2}
>\left(540\,\pi^2\right)^{2/3}\,P_\zeta^{2/3}\,T_{BBN}^{4/3}\sim
10^{-33},\label{c}
\end{equation}
where we have taken the scalar spectral index $n_s=1+(4m/9H_*)^2$
closed to one, and therefore $m\leq 0.1\,H_*$ (see
Ref.\cite{dimo}).

 We note here that if the curvaton decays before the
electroweak scale (since the baryogenesis is located below the
electroweak scale)  and one needs that the reheating temperature
should satisfy   $T_{reh}>T_{ew}$, where $T_{ew}$ is the
electroweak temperature. This inequality is a much stronger bound
than  $T_{reh}>T_{BBN}$ i.e., $T_{reh}>T_{ew}>T_{BBN}$. In this
way, we replace  $T_{BBN}$ by $T_{ew}$ in Eq.(\ref{c}), and thus,
we get the constraint $H_*^2>10^{-26}$. Here, we have used that
$T_{ew}\sim10^{-17}$.

 Also, we noted that if the decay rate is of gravitational
strength, then $\Gamma_\sigma\sim m^3$ (see
Refs.\cite{M4,M5,S1,S2}) and Eq.(\ref{ww}) becomes
\begin{equation}
m^2 < \frac{\,H_0^2}{54\pi^2\,P_\zeta}\,\left[ B\,N_* + \,\phi_e
^{\lambda\gamma} \right]^{\frac{2(\lambda
-1)}{\lambda}}\,\exp\left[ -\beta \left( B\,N_* + \phi_e
^{\lambda\gamma} \right)^{1/\lambda} \right].\label{wwq}
\end{equation}
In the same way,  Eq.(\ref{ew}) is now written as
\begin{eqnarray}
m^2\;<\,\frac{2}{3}\;\times10^{-4}\,\beta^2\,\gamma^2\;\;\;
\left[\frac{N_*}{A}\,\left(\frac{2(A\lambda)^{1/2}}{\lambda+1}\right)^{(\lambda+1)/2\lambda}+
\left(\frac{\beta\gamma}{\sqrt{2}}\right)^{(\lambda-1)/\lambda}\right]^{(1-\lambda)/\lambda}\,\,.\label{ew1}
\end{eqnarray}
These expressions gives an upper limits on the curvaton mass $m$,
when the constraints of the gravitational strength are taken into
account.
%Now we considered the special case in which we fixed $\lambda=5$,
%$A= 10^{-6}$ and $N_*=60$. In this special case we obtained from
%expression (\ref{ew1}) that $m^2<6\times 10^{-8}$.

%the typical scale for modular inflation is $V_*^{1/4}\sim
%\sqrt{m_{3/2}\,m_p}\sim 10^{10.5}$ GeV, where in the electroweak
%scale $m_{3/2}\sim 1$ TeV (see Refs.\cite{S1,S2}). In this case
%the energy becomes
%\begin{equation}
%H_*^2\simeq\frac{V_*}{3}\gtrsim 10^{-32}.\label{elec}
%\end{equation}
%Here, we have used that $8\pi/m_p^2=1$. We clearly see that the
%constraint given by Eq.(\ref{c}), is very similar to that
%expressed by Eq.(\ref{elec}) when the curvaton decays before the
%electroweak scale.

%Note also that this
%constraint provides a lower limit for the parameters $A$ and
%$\beta$. Alternatively,
% following the same line of reasoning of Ref. \cite{BuDi}, we could write an
%upper limit for the Hubble parameter $H_*$, which satisfies the
%inequality $H_*\leq 10^{-5}$.

%%%%%%%%%%%%%%%%%%%%%%%%%%%Chilean people's figures%%%%%%%%%%%%%%%%%%%%%%%%%%%%%%%%%%%%%%%%

%\begin{figure}[th]
%\includegraphics[width=10.0in,angle=0,clip=true]{fig.eps}
%\vspace{-9 cm}\caption{ Contour plot for the curvaton field,
%$\sigma_*^2$, as a function of the parameters $\lambda$ and $A$,
%according to Eq. (\ref{pafter}). Here, we have taken the values
%$N_*=60$ and $P_\zeta=2.9\times 10^{-9}$. \label{cou}}
%\end{figure}

\begin{figure}[th]
\includegraphics[width=5.0in,angle=0,clip=true]{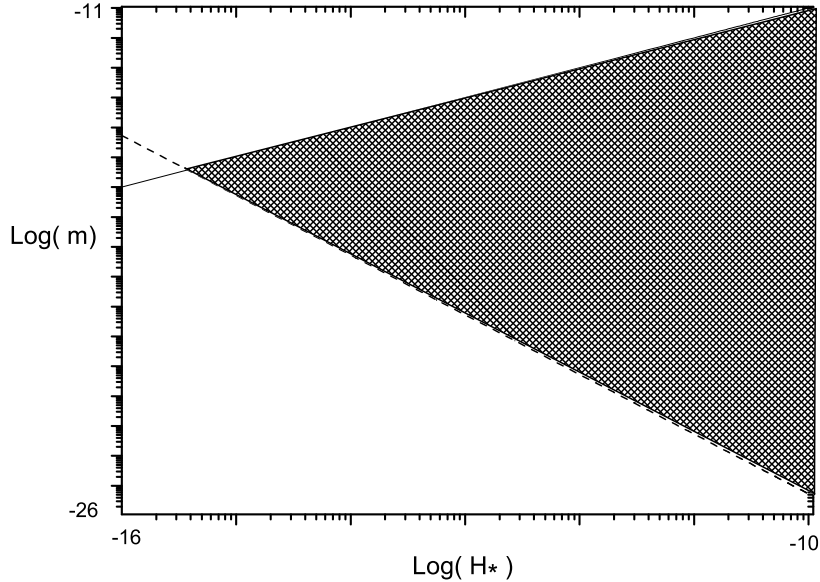}
\caption{ This plot shows the dependence of  the  curvaton mass,
$m$, as a function of the Hubble parameter, $H_*$, according to
Eqs. (\ref{gamm1}), (\ref{pafter})  (dashed line), $m\leqslant
0.1H_*$ (solid line) and the inequality (\ref{c}). The shadow zone
shows the allowed values of $m$ and $H_*$ parameters.\label{cou}}
\end{figure}

%%%%%%%%%%%%%%%%%%%%%%%%%%%%%%%%%%%%%%%%%%%%%%%%%%%%%%%%%%%%%%%%%%%%%%%%

%\vspace{0.5cm}

%\noindent
%Estimados, ustedes ya tenia una figura hecha en Mathematica

%%%%%%%%%%%%%%%%%%%%%%%%%%%%%%%%%%%%%%%%%%%%
 In  Fig. \ref{cou} we show the dependence of  the curvaton
mass, $m$ as a function of the Hubble parameter, $H_*$, according
to Eqs.(\ref{gamm1}), (\ref{pafter}), (\ref{c}) and $m\leqslant
0.1H_*$. We have taken $P_\zeta=2.4\times10^{-9}$.

Following the analysis done in Ref.\cite{A2}, we can obtain the
behavior of the parameters $\gamma$ and $\alpha$ in all regions of
the parameter space. This behavior is given   in terms of the
parameters $\lambda$, $A$ and $m$. In this aspect, we can
determine the parameters in which the logamediate inflation
together with the  curvaton scenarios work. It is known that when
$0\leq\gamma\leq1$, inflation is generic for late times, and for
any value of $\alpha$, and, in our case, it reads as follow,
$\lambda \geq 1$ and $\alpha \geq 0$. For the case, when
$\gamma>1$ and $\alpha\neq 0$, the models are noninflationary,
because $\lambda < 1$. When $\gamma<0$, $\alpha\geq 0$ or
$\alpha<0$, this model does not work since $\lambda < -1$. For
$\gamma>1$, $\alpha \neq 0$, or $\alpha =0$, these models  never
can inflate, due to $\lambda < 1$ .
 In particular, if $\gamma=0$ (or equivalently $\lambda \rightarrow
 \infty$), this model does not work since from Eq.
 (\ref{ew}) $\Gamma_{\sigma}<0$. The case when
$\alpha=0$  de Sitter solution is obtained and $\gamma=1$
($\lambda=1$), and from Eq.(\ref{ew1}) and the analysis done in
Ref. \cite{R12}, we obtain the following constraint
$1<A<10^{-4}/m^2$ in order to have a working model. For the cases
$\gamma>1$ ($\lambda>1$) and $\alpha>0$, from Eq. (\ref{ew1}) we
find that our model works for  $m^2<10^{-5}$ and $A>0$.

%In  Fig. \ref{cou} we have plotted the contour curves that
%correspond  to  different values of the curvaton field,
%$\sigma_*^2$, for a set of the space of the $\lambda$ and $A$
%parameters,  according to Eq.(\ref{pafter}). Here, we have taken
%$N_*=60$ and $P_\zeta=2.4\times10^{-9}$.
% A similar
%graph is obtained when the upper limit, $H_* \leq 10^{-5}$, is
%used, except that the contour lines get bigger values for the
%$N_*$ parameter (see Fig.\ref{cou2}).
%\begin{equation}
%{\frac {1}{96}}\,{A}^{2}{l}^{2} \left( {\frac {N}{A}}+ \left(
%1/2\,\left( l+1 \right) ^{2} \left( \left( 1/4\,{\frac { \left(
%l+1\right) \sqrt {2}}{\sqrt {Al}}} \right) ^{2\, \left( l+1
%\right) ^{-1}} \right) ^{-2} \right) ^{1/4\,l \left( l+1 \right)
%\left( 2\,\left( l+1 \right) ^{-1}-1 \right) ^{-1}} \right)
%^{2\,{\frac {l-1}{l}}}{\pi }^{-2}{P}^{-1} \left( {{\rm e}^{2\,
%\left( {\frac {N}{A}}+\left( 1/2\, \left( l+1 \right) ^{2} \left(
%\left( 1/4\,{\frac {\left( l+1 \right) \sqrt {2}}{\sqrt {Al}}}
%\right) ^{2\, \left( l+1\right) ^{-1}} \right) ^{-2} \right)
%^{1/4\,l \left( l+1 \right) \left( 2\, \left( l+1 \right) ^{-1}-1
%\right) ^{-1}} \right) ^{{l}^{-1}}}} \right) ^{-1}
%\end{equation}

\section{Curvaton Decay Before Domination\label{sectv}}

On the other hand, considering that the curvaton $\sigma$ decays
before it dominates the expansion (which we called the second
scenario) and  additionally, the mass of the curvaton is
non-negligible when compared with the Hubble expansion rate $H$,
i.e., $m\sim H$ and if the curvaton field decays at a time when
$\Gamma_\sigma =H(a_d)=H_d$, where "d" stands for quantities at
the time when the curvaton decays, we get that

\begin{equation}
\Gamma_\sigma=\;H_d=\;H_k\;\left(\frac{a_k}{a_d}\right)^3,
\label{Gamm}
\end{equation}
where Eq.(\ref{h}) is used.

If we allow the decaying of the curvaton   after its mass becomes
important, i.e. $\Gamma_\sigma<m$, and before that the curvaton
dominates the cosmological  expansion (i.e.,
$\Gamma_\sigma>H_{eq}$), we may write the constraint

\begin{equation}
\frac{\sigma_*^2}{6}<\frac{\Gamma_\sigma}{m}<1, \label{gamm2}
\end{equation}
which is similar to  that described in Ref.\cite{R21}.

In this scenario, the curvaton decays at the time when
$\rho_\sigma<\rho_\phi$. Denoting the  parameter $r_d$ as the
ratio between the curvaton and the inflaton energy densities,
evaluated at the time in which the curvaton decay occurs, i.e. at
$a=a_d$,  the  parameter $P_\zeta$, becomes given by
\cite{ref1u,L1L2}

\begin{equation}
P_\zeta\simeq \frac{r_d^2}{16\pi^2}\frac{H_*^2}{\sigma_*^2}.
\label{pbefore}
\end{equation}

Since $r_d=\left.\frac{\rho_\sigma}{\rho_\phi}\right|_{a=a_d} $,
we get that $ r_d=\frac{m^2\;\sigma_*^2\; a_m^3
\;a_{d}^3}{6\;H_k^2\; a_k^6} $, where we have used
$\rho_\sigma(a)= \frac{m^2\sigma_*^2}{2}\left (
\frac{a_k}{a}\right )^3$ and $\rho_\phi(a)=\rho_\phi^k\left (
\frac{a_k}{a}\right )^6$. Using Eqs.(\ref{mh}) and (\ref{Gamm}) we
obtain
\begin{eqnarray}
r_d=\;\frac{m\;\sigma_*^2}{6\;\Gamma_\sigma}\label{rd}.
\end{eqnarray}
 The parameter $r_d$ is related to two other observable, the
amount of non-Gaussianity, that is conventionally specified by a
number $f_{NL}$ (NL meaning "nonlinear") \cite{S3} and the
parameter of the isocurvature amplitude (or the ratio the
isocurvature and adiabatic amplitudes at the pivot scale)
\cite{S4}. Following, Ref.\cite{L1L2}  the parameter $f_{NL}$,
becomes of the order of $f_{NL}\simeq\frac{5}{4\,r_d}$, where this
expression is only valid for high value of $f_{NL}$, which
dominates over the intrinsic non-Gaussianity ( see Ref.\cite{S5}
for the second order perturbations).  From the  observational data
$f_{NL}<100$, therefore the parameter $r_d$ satisfies  $r_d>
0.01$\cite{M2,L1L2}.

 %For $f_{NL}\sim100$, it makes
%the non-gaussianity signal to contribute with just 0.1$\%$ of the
%total curvature perturbation \cite{M2,L1L2}. This gives that $r_d$
%is of the order of $r_d\sim 0.01$, which validates that the
%curvaton decays during radiation regime.

From Eqs. (\ref{pbefore}) and (\ref{rd}) we find that $
\sigma_*^2=576\;\pi^2\;\frac{P_\zeta}{m^2}
\frac{\Gamma_\sigma^2}{H_*^2}\,$ and using that
\begin{equation}
 H^2_*=H_0^2\,\left[ B\,N_* +
\,(\beta^2\gamma^2/2) ^{\lambda/(\lambda-1)}
\right]^{\frac{2(\lambda -1)}{\lambda}}\,\exp\left[ -\beta \left(
B\,N_* + (\beta^2\gamma^2/2) ^{\lambda/(\lambda-1)}
\right)^{1/\lambda} \right] \label{h2},
\end{equation}
we get
\begin{equation}
\sigma_*^2=576 \pi^2\frac{P_\zeta\;\Gamma_\sigma^2}{m^2\,H_0^2}
\left[ B\,N_* + \,(\beta^2\gamma^2/2) ^{\lambda/(\lambda-1)}
\right]^{\frac{2( 1-\lambda)}{\lambda}}\,\exp\left[ \beta \left(
B\,N_* + (\beta^2\gamma^2/2) ^{\lambda/(\lambda-1)}
\right)^{1/\lambda} \right] .\label{sg}
\end{equation}
Thus, expressions (\ref{gamm2}) and (\ref{sg}) becomes useful for
obtaining the following inequality for the decay parameter
$\Gamma_\sigma$
\begin{equation}
\Gamma_\sigma< \frac{m\,H_0^2}{576\pi \,P_\zeta}\,\,\left[ B\,N_*
+ \,(\beta^2\gamma^2/2) ^{\lambda/(\lambda-1)}
\right]^{\frac{2(\lambda -1)}{\lambda}}\,\exp\left[ -\beta \left(
B\,N_* + (\beta^2\gamma^2/2) ^{\lambda/(\lambda-1)}
\right)^{1/\lambda} \right] .
%\frac{1}{18\pi^2}\frac{m\;A^2}{P_\zeta(\beta+4)^2}
%\left[\frac{2\,A}{\beta+1-2\,N_*}\right]^{\beta/2}
\label{37}
\end{equation}

%We see that this inequality for $\Gamma_\sigma$ depends on the
%free parameters, $A$ and $\beta$, characteristic of the
%intermediate inflationary universe model.

We also derive a new constraint for the parameters $A$ and
$\lambda$ characteristic of the logamediate inflationary universe
model, by using the  BBN temperature $T_{BBN}$. Since, the
reheating temperature satisfies the bound $T_{reh}>T_{BBN}$, with
$\Gamma_\sigma\,>\,T^2_{BBN}$ we get
\begin{eqnarray}
%16 \left( \frac{A}{4+\beta}\right)^2 \left
%[\frac{2\,A}{\beta+1-2\,N_*} \right]^{\beta/2}
H_0^2\,\left[ B\,N_* + \,(\beta^2\gamma^2/2)
^{\lambda/(\lambda-1)} \right]^{\frac{2(\lambda
-1)}{\lambda}}\,\exp\left[ -\beta \left( B\,N_* +
(\beta^2\gamma^2/2) ^{\lambda/(\lambda-1)} \right)^{1/\lambda}
\right]
 \nonumber
 \\
>\left(960\,\pi^2\right)^{2/3}\,P_\zeta^{2/3}\,T_{BBN}^{4/3}\sim
10^{-33}.\label{c3}
\end{eqnarray}
Here, we have used that $m\simeq H_*/10$ see Ref.\cite{dimo} and
Eqs. (\ref{gamm2}), (\ref{rd}) and (\ref{h2}).

%Note that this inequality   is similar to that obtained when the
%curvaton field decays before domination, given  by Eq.(\ref{c}).

From the hypothesis that
$P_{\zeta_{\phi}}\lesssim\,0.0001\,P_\zeta$,
 we can set a new constraint for the decay
parameter $\Gamma_\sigma$ given by
$$
\Gamma_\sigma\lesssim\frac{10^{-2}}{12}\,\beta\,\gamma\,m\,\sigma_*^2\,\phi_*^{(1-\gamma)},
$$
or equivalently,
\begin{eqnarray}
\Gamma_\sigma\lesssim\,8\times10^{-4}\,\beta\,\gamma
\left[\frac{N_*}{A}\,\left(\frac{2(A\lambda)^{1/2}}{\lambda+1}\right)^{(\lambda+1)/2\lambda}+
\left(\frac{\beta\gamma}{\sqrt{2}}\right)^{(\lambda-1)/\lambda}\right]^{(\lambda-1)/2\lambda}\,\;\;m\,\sigma_*^2.
\end{eqnarray}
Here, we have used Eqs. (\ref{nn}), (\ref{pbefore}) and
(\ref{rd}).

 On the other hand,  if the decay rate is of gravitational
strength, then $\Gamma_\sigma\sim m^3$ and the Eq.(\ref{37}) is
given by

\begin{equation}
m^2< \frac{\,H_0^2}{576\pi \,P_\zeta}\,\,\left[ B\,N_* +
\,(\beta^2\gamma^2/2) ^{\lambda/(\lambda-1)}
\right]^{\frac{2(\lambda -1)}{\lambda}}\,\exp\left[ -\beta \left(
B\,N_* + (\beta^2\gamma^2/2) ^{\lambda/(\lambda-1)}
\right)^{1/\lambda} \right] ,
%\frac{1}{18\pi^2}\frac{m\;A^2}{P_\zeta(\beta+4)^2}
%\left[\frac{2\,A}{\beta+1-2\,N_*}\right]^{\beta/2}
\label{378}
\end{equation}
and also the curvaton mass should obey the constraint
$m^2\lesssim\frac{10^{-2}}{12}\,\beta\,\gamma\,\,\sigma_*^2\,\phi_*^{(1-\gamma)}$,
where it was consider that
$P_{\zeta_{\phi}}\lesssim\,0.0001\,P_\zeta$ as before.

\section{Gravitational waves\label{sectgw}}
Another set of bounds is due to the possible overproduction of the
gravitational waves due to inflation. The corresponding
gravitational wave amplitude   can be written as
\begin{equation}
h_{GW}\simeq\,C_1\,H_*,\label{po}
\end{equation}
where the constant $C_1\approx 10^{-5}$ \cite{Staro2}.

We may write  the gravitational  wave
 amplitude as function of the numbers of e-folds of inflation,
 i.e.
\begin{equation}
h_{GW}^2\simeq\,\;C_1^2\;\left[ \frac{N_{*}}{A} + B\left(
\frac{2}{\beta^2 \gamma^2}\right)
 ^{\frac{\gamma}{\gamma(2\gamma -1)}}   \right]^{\frac{2(\lambda -1)}{\lambda}}
 \,\exp \left\lbrace - \frac{2}{\gamma} \left[ \frac{N_{*}}{A} + B\left( \frac{2}{\beta^2
 \gamma^2}\right)^{\frac{\gamma}{\gamma(2\gamma -1)}}   \right]
 \right\rbrace
  ,\label{gw}
\end{equation}
where we have used Eq.(\ref{h2}).

After inflation the inflaton field follows an equation of state
which is almost stiff and  the spectrum of relic gravitons
presents a characteristic in which the slope grows with the
frequency (spike) for models that reenter the horizon during this
epoch. This means that at high frequencies the spectrum forms a
spike instead of being  flat, as in the case of radiation
dominated universe\cite{Gi}. Therefore, high frequency gravitons
reentering the horizon during the kinetic epoch may disrupt BBN by
increasing the Hubble parameter. This problem can be  avoided if
the following constraint on the density fraction of the
gravitational wave is required \cite{R19} (see also
Ref.\cite{dimo})
\begin{equation}
I\equiv\,h^2\,\int_{k_{BBN}}^{k_{*}}\;\Omega_{GW}(k)\,d\ln\,k\simeq\,2\,
h^2\,\epsilon\,\Omega_\gamma(k_0)\,
h_{GW}^2\left(\frac{H_*}{ \widetilde{H}}\right)^{2/3}\leq\,2\times
10^{-6},\label{I}
\end{equation}
where $\Omega_{GW}(k)$ is the density fraction of the
gravitational wave with physical momentum $k$, $k_{BBN}$ is the
physical momentum  corresponding to the horizon at BBN,
$\Omega_\gamma(k_0)=2.6\times 10^{-5}h^{-2}$ is the density
fraction of the radiation at present on horizon scales. Here,
 $h=0.73$ is the Hubble constant in
which $H_0$ is in units of 100 km/sec/Mpc and $\epsilon\sim
10^{-2}$.  The parameter $\widetilde{H}$ represents either
$\widetilde{H}=H_{eq}$, when the curvaton decays after domination,
or $\widetilde{H}=H_{d}$, if the curvaton decays before
domination.

For the first  scenario,  the constraint on the density fraction
of the gravitational wave, expressed by Eq.(\ref{I}), becomes
\begin{equation}
\frac{m}{\sigma_*^2}\gtrsim\,\left(\frac{\,P_{\zeta}}{4\times10^{5}}\right)^2\sim
10^{-28},\label{II}
\end{equation}
where we have used  expressions (\ref{heq}), (\ref{po}) and
$C_1\sim10^{-5}$. From Eqs.(\ref{gamm1}) and (\ref{II}) we
obtained that $m>10^{-34}$ and
$10^{-40}/m<\sigma_*^2\lesssim10^{28}\,m$.

For the second scenario, the constraint on the density fraction of
the gravitational wave given by Eq.(\ref{I}), becomes
\begin{equation}
\frac{m^2\;\sigma_*^2}{\Gamma_\sigma^{1/4}}\gtrsim\,6\times10^{-5}\,P_{\zeta}\sim
10^{-13},\label{I1}
\end{equation}
where we have used  Eqs.(\ref{Gamm}) and (\ref{pbefore}). From
Eqs.(\ref{gamm2}) and (\ref{I1}), we may write the inequality for
the parameter given by $\Gamma_\sigma>10^{-19}\,m^{-4/3}$.

\section{Conclusions \label{conclu}}
We have studied in detail the curvaton mechanism into the NO
inflationary logamediate model. The curvaton scenario is
responsible for reheating the Universe as well as for the
curvature perturbations.

%We have introduced the curvaton field in the logamediate
%inflationary universe model.
In describing the curvaton reheating   we have considered two
possible scenarios. In the first one, the curvaton dominates the
universe after it decays and thus we have obtained  the upper
limit for $\Gamma_\sigma$ expressed by Eq.(\ref{ww}). In the
second scenario the curvaton decays before domination. Here, we
have also found a constraint for the values of $\Gamma_\sigma$
which is represented by Eq.(\ref{37}).

%In the context of the curvaton scenario, reheating does occur at
%the time when the curvaton decays, but only in the period when the
%curvaton dominates. In contrast, if the curvaton decays before its
%density dominates the universe, reheating occurs when the
%radiation due to the curvaton decay manages to dominate the
%universe.

During the  scenario in which the curvaton decays after its
dominates, our computations allow us to get  the reheating
temperature $T_{rh}\propto\Gamma_{\sigma}^{1/2}$ as hight as
$10^{-12}$ (in units of $m_p$). Here, we have used Eq. (\ref{ww}),
with $m\sim 10^{-8}$, $N_*=60$, $P_\zeta=2.4\times 10^{-9}$,
$\lambda=5$
 and $A\simeq10^{-6}$ (see Ref.\cite{R12} for the values of $\lambda$ and
 $A$). In particular, for $\lambda=10$ and $A\simeq10^{-15}$ we get
 that the reheating temperature is of the order of $10^{-16}$.
 In the case when $\lambda=2$ and $A\simeq
2\times10^{-2}$ the reheating temperature is of the order of
$T_{rh}\sim 10^{-21}$. If we consider the constraint from
gravitational wave, we find that the inequalities  for the scalar
field $\sigma_*$ at the moment when the  cosmological scales exit
the horizon becomes $10^{-16}<\sigma_*<\sqrt{6}$.

In the second scenario, we could estimate the reheating
temperature to be  of the order of $\sim 10^{-13}$ as an upper
limit from Eq.(\ref{gamm2}).  Here, we have used  $m\sim 10^{-8}$,
$N_*=60$, $P_\zeta=10^{-10}$, $\lambda=5$ and $A\simeq10^{-6}$. In
particular, for $\lambda=10$ and $A\simeq10^{-15}$ we estimate
$T_{rh}\sim 10^{-17}$. For the values $\lambda=2$ and $A\simeq
2\times10^{-2}$ the reheating temperature is of the order of
$T_{rh}\sim 10^{-22}$. From the constraint of the gravitational
wave we have obtained that $T_{rh}>10^{-13}$.   Note that the
value of this  temperature does not agree with the previous value,
this is due the fact that its value is obtained from different
cosmological constraints.  However, we have obtained values for
the reheating temperature $T_{rh}$ which are in good
 agreement with those values
reported previously  in Refs.\cite{dimo,cdch}, which seriously
challenges gravitino constraints, where the reheating temperature
becomes of the order  of  $T_{rh}\sim 10^{-9}$ \cite{Elis} .

\begin{acknowledgments}
SdC wishes to thank John Barrow for calling attention to the
logamediate inflationary universe models. This work was supported
by Comision Nacional de Ciencias y Tecnolog\'{\i}a through
FONDECYT  Grants 1070306 (SdC), 1090613 (RH, SdC and JS) and
11060515 (JS). CC and ER acknowledges "FONDECYT-Concurso incentivo
a la cooperaci\'on internacional" No. 7080205, and are grateful to
the Instituto de F\'{\i}sica for warm hospitality. Also this work
was partially supported by DI-PUCV 2009. CC and ER acknowledge
support from the grant, PROMEP PIFI 3.3  (C.A. INVEST. Y ENSEÑ. DE
LA FÍS.), and CC acknowledges to grant PROMEP 103.5/08/3228.

\end{acknowledgments}

%\\\\\\\\\\\\\\\\\\\\\\\\\\\\\\\\\\\\\\\\\\\\\\\\\\\\\\\\\\\\\\\\\\\\\\\

\end{document}